\documentclass[aps,prb,superscriptaddress,twocolumn,10pt]{revtex4-1}
\usepackage{graphicx}
\usepackage{wasysym}
\usepackage{listings}
\usepackage{color}

\lstdefinelanguage[10]{Mathematica}[5.2]{Mathematica}%%
  {morekeywords={ImageAdjust,Image,ImageConvolve,GaussianMatrix,%
	 MaxDetect,CornerNeighbor,Binarize,ColorNegate,DiskMatrix,%
	DeleteSmallComponents,Dilation,MorphologicalComponents,%
	SelectComponents,GrowCutComponents,ImageSubtract,CornerNeighbors,%
	Colorize,ConstantArray,ComponentMeasurements,CrossMatrix,ImageCrop,%
	ImageData,Erosion,ImagePad,Parallelize,Normalize,NonlinearModelFit,%
	Quiet,FindPeaks},%
}

\begin{document}

\hyphenation{To-po-gra-phy}

\lstset{language=[10]Mathematica}

\newcommand{\didv}{\ensuremath{\mathrm{d}I/\mathrm{d}V}}
\newcommand{\iv}{\ensuremath{I(V)}}
\newcommand{\mV}{\mathrm{mV}}
\newcommand{\uV}{\ensuremath{\mu\mathrm{V}}}
\newcommand{\Vb}{\ensuremath{V_\mathrm{bias}}}
\newcommand{\Qf}{\ensuremath{Q_\mathrm{f}}}
\newcommand{\qe}{\ensuremath{e}}
\newcommand{\sli}{$S_\mathrm{LI}$}
\newcommand{\siv}{$S_\mathrm{LI,IV}$}
\newcommand{\sfu}{$S_\mathrm{LI,FV}$}
\newcommand{\CO}{(Color Online)\ }
\newcommand{\RA}{\textsf{``R1''}}
\newcommand{\RB}{\textsf{``R2''}}
\newcommand{\ROI}{\textsf{``ROI''}}
\newcommand{\CB}[1]{\textsf{``CB{#1}''}}

\title{Coulomb blockade regions in sputter-deposited titanium nitride films}
\author{Michael Dreyer, Peng Xu, Kevin D. Osborn, R. E. Butera}
\email{dreyer@lps.umd.edu}
\affiliation{Department of Physics, University of Maryland, College Park, MD 20740, USA.}
\affiliation{Laboratory for Physical Sciences, 8050 Greenmead Drive, College Park, MD 20740, USA.}

\date{\today}

\begin{abstract}
We present topographic and spectroscopic scanning tunneling microscopy measurements taken on a 21~nm thick TiN film at a temperature of $4.2$~K -- above the superconducting transition temperature ($T_\mathrm{c}=3.8$~K) of the sample. The film was polycrystalline with crystallite diameters of $d\approx 19$~nm, consistent with other films prepared under similar conditions. The spectroscopic maps show on average a shallow V-shape around $\Vb=0$~V consistent with a sample near the Mott insulation transition. In selected regions on several samples we additionally observed signs of Coulomb blockade. The corresponding peak structures are typically asymmetric with respect to bias voltage indicating coupling to two very different tunneling barriers. Furthermore, the peak structures appear with constant peak-peak spacing which indicates quantum dot states within the Coulomb blockade island. In this paper we discuss one such Coulomb blockade area and its implications in detail.  
\end{abstract}

\maketitle

\section{Introduction}

Titanium nitride, as a superconducting film in devices, has the useful properties of high kinetic inductance and low microwave loss. The latter property is related to the TiN film quality as well as its good material-interface formation\cite{Vissers:2010} which aids low noise in astronomy detectors\cite{Leduc:2010,Diener2012} and high-coherence in quantum bits\cite{Chang:2013}. However, the former property is perhaps the most interesting as it brings new qualitative features to recent devices. Higher kinetic inductance generally appears with a lower $T_\mathrm{c}$ of the film and affects the minimum photon detection frequency in the so-called kinetic-inductance detectors. The high inductance can also enable intentional phase slips in a nanowire-based superconducting quantum bit\cite{Astafiev:2012}. High kinetic inductance nitrides have also been recently exploited in traveling-wave amplifiers\cite{Eom:2012} and for tunable coupling in resonators\cite{Bockstiegel:2016}.

Knowledge about the topographic and electronic structure is important for device property analysis. While there are many different growth techniques for TiN\cite{Vissers:2010,Jaim:2015,Ohya:2014,Leduc:2010,Torgovkin:2015} the films generally contain different crystalline orientations. The dominant orientation is correlated with resonator quality\cite{Vissers:2010,Jaim:2015,Ohya:2014} in contrast to BCS theory of a uniform film. Anomalous qualitative features in films including excess quasiparticles\cite{Driessen:2012} and quasiparticle relaxation\cite{Bueno:2014} are attributed to a short mean free path or to inhomogeneity of the nitride superconductor. In a small energy range an anomalous pseudo-gap feature has been observed above $T_\mathrm{c}$ related to unexpected pre-formed pairing\cite{Sacepe:2010}. The properties above $T_\mathrm{c}$ can be quite important since the kinetic inductance of TiN can be increased by proximity to a Mott-insulator non-Fermi-liquid state which suppresses the density of states at the Fermi energy\cite{Allmaier:2009}.

Here we discuss recent results obtained on sputter-prepared polycrystalline TiN using scanning tunneling microscopy (STM) and scanning tunneling spectroscopy (STS). The measurements were performed at a temperature of $4.2$~K. The superconducting transition temperature of the sample was $T_\mathrm{c}=3.8$~K. Hence, the samples were observed in the normal state. The main focus was to study the topographic and electronic structure as well as their homogeneity since either conceivably influence device performance and variability. 

In some regions the spectroscopic data show signs of ``naturally occurring'' Coulomb blockade (CB) as discussed below. Coulomb blockade in STM is usually observed when a part of the sample is coupled to the bulk of the sample by a high resistance channel similar in magnitude to the STM tunnel junction. With the STM tip acting as a second electrode, this part then behaves like the island of a fabricated single electron transistor (SET). The tip acts as the drain electrode as well as a moveable gate with a position dependent tip-island capacitance. The bulk of the sample acts as the source with a fixed capacitance and resistance value. Hence one of the telltale signs of CB in spectroscopic STM data is a systematic change of the energetic position of the CB peaks with the tip position limited to the area of the crystallite in question. Incidentally, a particle can be picked up by the tip which can lead to CB being observed throughout the image area. However, here we focus on the case of local regions of CB where the source is clearly on the sample surface. In careful examination of 47 STS maps taken over four different samples we found local CB in nanometer sized areas in 12 maps representing all samples. Considering the total area scanned in the STS maps of $A_\mathrm{STS}\approx 656\times 656$~nm$^2$, the presence of CB is a relatively common occurrence on our TiN samples.

\section{Sample preparation and measurement}

The titanium nitride samples were prepared in an external DC magnetron sputtering system on a high resistivity ($\rho>20000$~$\mathrm{\mu\Omega\cdot cm}$, float-zone, 3 inch diameter) silicon (100) substrate. The silicon wafer was cleaned with a 49\% HF solution for 30~seconds to remove the native oxide. The substrate was then placed in an O$_2$ plasma for 30~seconds before being put into the sputtering system. During the TiN deposition, a constant pressure of 3.5~mTorr was maintained in the presence of Ar and N$_2$, flowed at $\dot{m}_\mathrm{Ar}=15$~SCCM and $\dot{m}_\mathrm{N_2}=10$~SCCM, respectively. A power of 400~W was applied to a 3~inch Ti target for the film depositions. Pre-sputtering was performed for 1~minute without RF-bias, and an additional 30~seconds with the RF power applied to the substrate prior to opening the substrate shutter\cite{Jaim:2015}. The substrate temperature was kept at 500~$^\circ$C. 

The wafer was cut into $1.0\times0.4$ cm$^2$ samples, which were mounted on an STM sample holder and sonicated in isopropanol to remove accumulated small particles due to exposure to air. After introducing them into the STM system\cite{our_4K_system} the samples were further cleaned by heating to $T\sim 400$~$^{\circ}$C for $30-60$ minutes. During the first $20-30$ minutes of the heating cycle the samples were also Ar$^+$-etched at $450-600$~eV ion energy and an ion flux of $j=10^{13}$~s$^{-1}$cm$^{-2}$. This preparation was found sufficient to remove insulating surface contamination without drastically changing the surface structure. The samples were then put into the STM and measured at 4.2~K. We measured the topography and took spectroscopic maps by simultaneously acquiring $I(V)$ and \didv\ curves at each point within the scan area. The \didv\ curves were measured by adding a modulation to the bias voltage and detecting the signal in the tunneling current using a lock-in amplifier. The modulation frequency was $f_\mathrm{LI}=1973$~Hz and the amplitude was chosen to match the respective voltage resolution of the $I(V)$ curve to maximize the signal to noise ratio while avoiding averaging over adjacent voltage steps.

We used etched tungsten wires as STM tips. The tips were cleaned by field emission on a gold (111) single crystal. The success was verified on the gold crystal by measuring a straight, metallic \iv\ curve and a work function of at least 4~eV in current vs.\ distance curves. 

\section{Results and discussion}

\subsection{Surface structure}

\begin{figure}%
\includegraphics[width=.475\columnwidth]{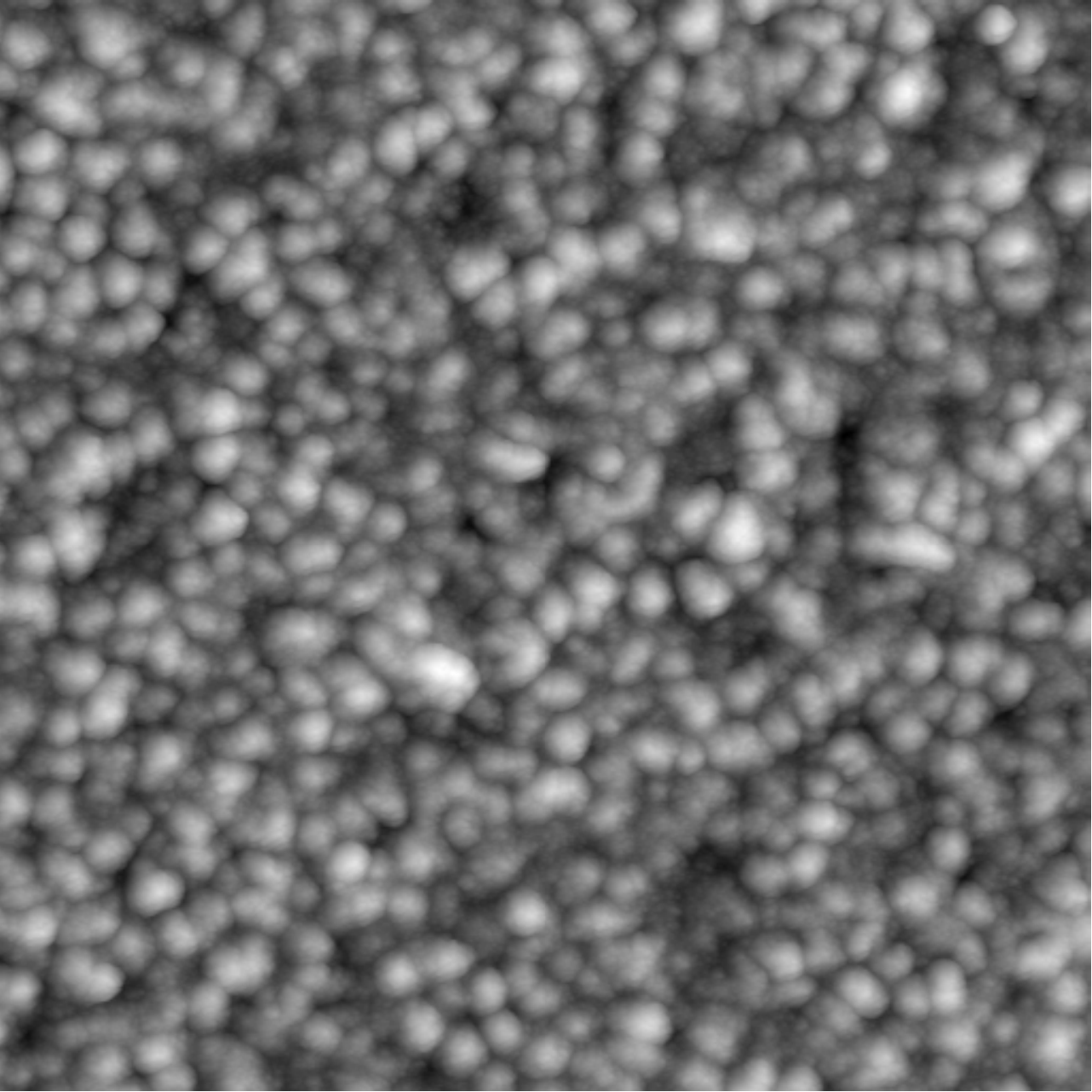}\hfill%
\includegraphics[width=.45\columnwidth]{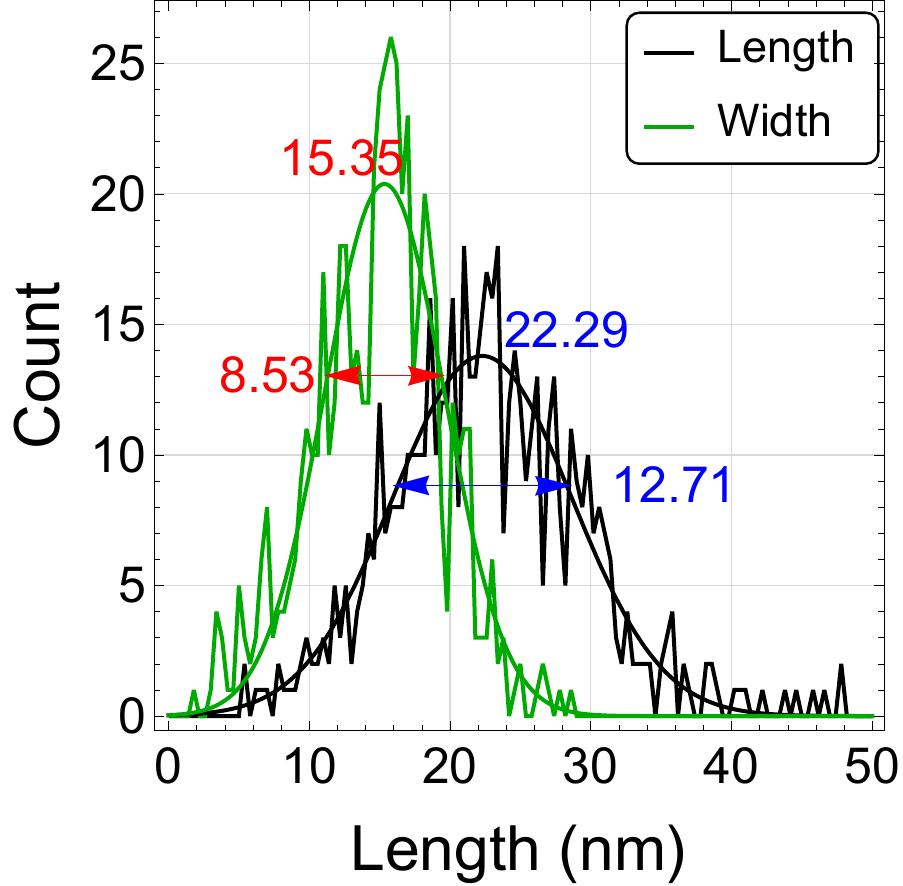}%
\caption{\CO Topographic image ($400\times 400\ \mathrm{nm}^2$, $\Vb=800$~mV, $I_\mathrm{t}=75$~pA, $\Delta z=7.7$~nm, $z_\mathrm{rms}=1.4$~nm) and histogram of particle length (22~nm) and width (15~nm) based on fitting individual crystallites to an equal-area ellipse. The histograms are fit to a Gaussian distribution to extract center position and distribution width.}%
\label{topography}%
\end{figure}

A typical topographic image of an area of $400\times 400$~nm$^2$ is shown in Fig.~\ref{topography}. The films are polycrystalline with a typical diameter of $d=18.8\pm 4.8$~nm. The grain size distribution was determined by semi-automatically segmenting the image into individual grains using Mathematica\cite{Mathematica}. The algorithm used produces better results in separating crystallites than a simple watershed algorithm\cite{Watershed}. The individual segments are then fit to an ellipse of equal area. A histogram for the long and and short axis, respectively, are shown in Fig.~\ref{topography}. The means are $d_\mathrm{long}=22.3\pm 6.4$~nm and $d_\mathrm{short}=15.4\pm 4.2$~nm. The direction of the elongation however lies mostly in the x-direction and is thus at least in part due to piezo hysteresis. In atomically resolved low-temperature images at scan frequencies of 1 Hz the distortion amounts to approximately 20~\% but, of course, varies with scan size and scan speed.

\subsection{Spectroscopic data}

\begin{figure}%
\includegraphics[height=.45\columnwidth]{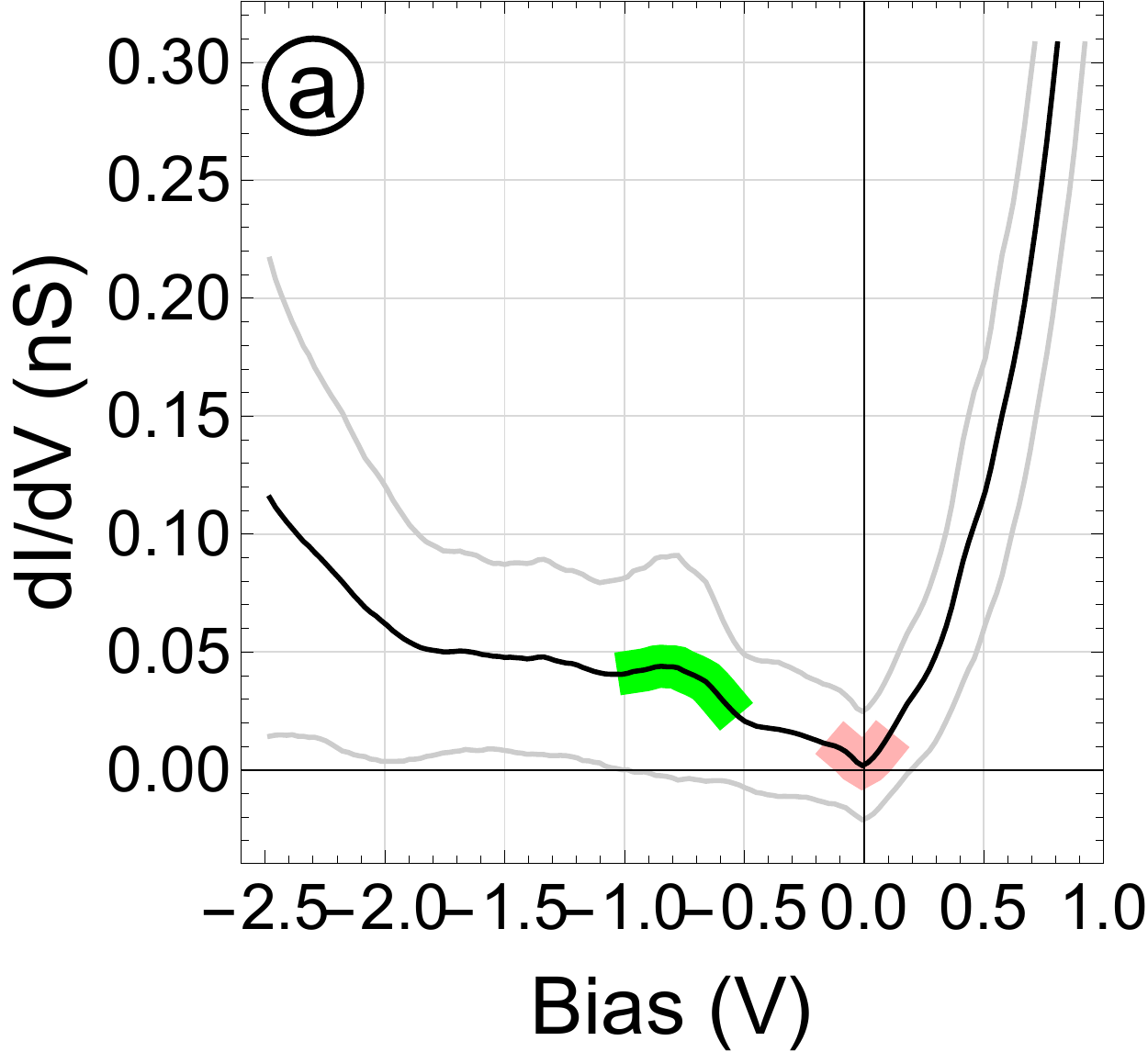}\hfill%
\includegraphics[height=.45\columnwidth]{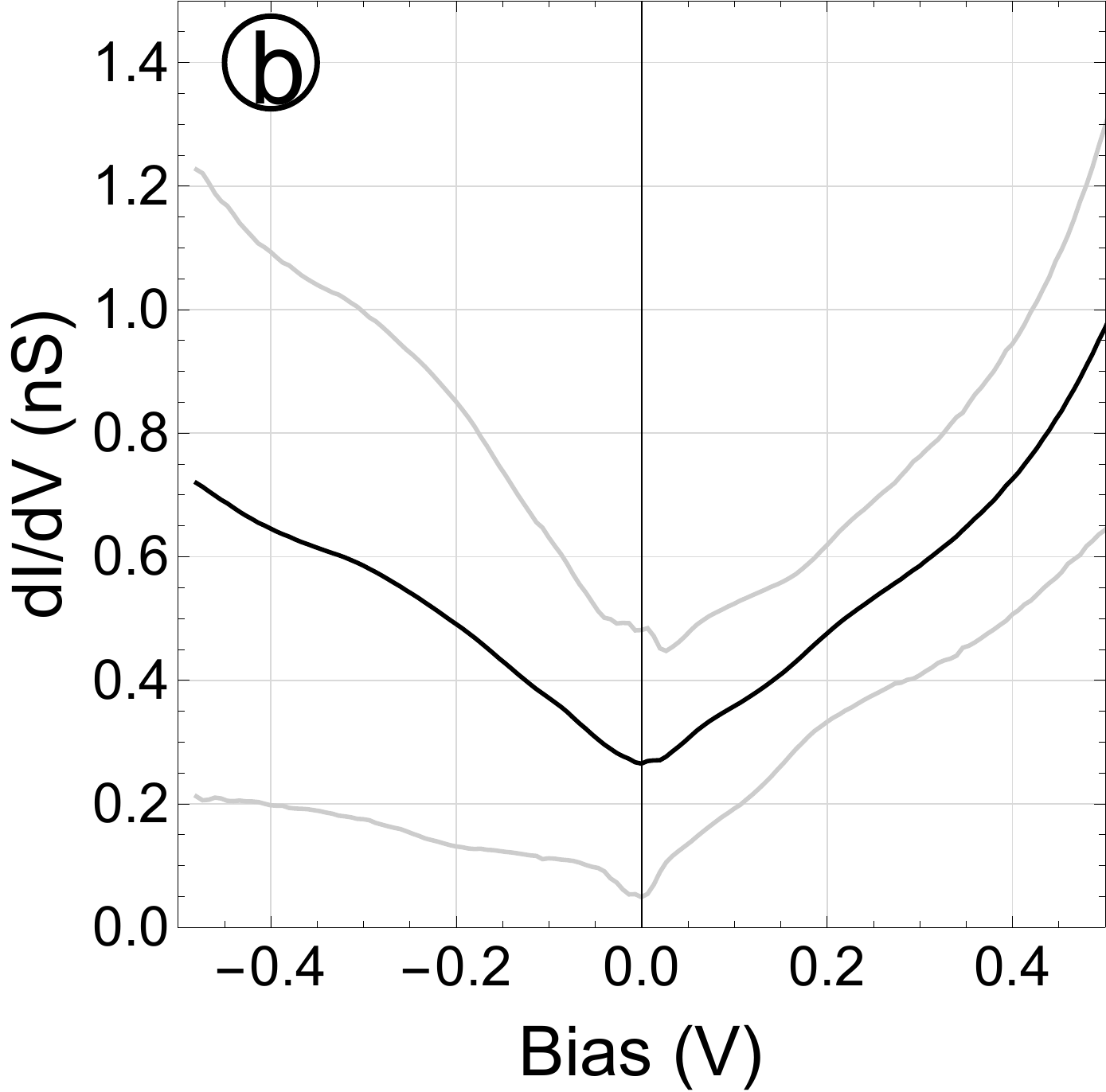}%
\caption{\CO (a) Large voltage range averaged \didv\ spectra showing a V-shaped gap, small sup-gap (red) and a peak like structure at 0.75~V (green). (b) At smaller voltage range a V-shaped gap with a small dip at 0 V remains. The gray lines mark the standard deviation of the numerical average at each bias voltage step.}\label{lrg_sts}%
\end{figure}

Most of the STS maps were taken in a voltage range of $\Vb=\pm 0.5$~V. While individual spectra may vary strongly from point to point, the average shows a distinct V-shape with a clear dip at $\Vb=0$~V. On a larger voltage scale the averaged spectra show a steep rise at positive bias and a peak at $\Vb=-0.75$ V. STS examples measured in two different voltage ranges and sample areas are shown in Fig.~\ref{lrg_sts}. The overall asymmetric structure qualitatively agrees with the calculated density of states which identifies a V-shape gap in an energy range of $\pm 1$ eV as the onset of Mott insulation\cite{Allmaier:2009}. We found additional local peak structures at or near zero bias as well as regions of Coulomb blockade in several areas on all samples investigated. The peaks might be gap states in the Mott insulator and will be discussed elsewhere\cite{md_TiNPeaks}. The CB states indicate that some of the crystallites or parts thereof are electronically only weakly coupled to the rest of the film network.  Here we focus on one of the more striking examples of CB which also shows internal quantum dot states. While we found regions showing CB on all samples studied at 4.2~K internal quantum dot states were a rare occurrence. In the context of superconductive devices CB islands could act as sources or traps for quasiparticles\cite{PhysRevB.72.014517,PhysRevB.76.069902} and therefore influence the coherence and performance.

\subsection{Coulomb blockade}

% 13-31,222
% substructure: QD
% first peak position: CB, other peaks: QD

%\begin{figure}%
%\includegraphics[width=0.49\columnwidth]{sts6_toposource.pdf}\hfill%
%\includegraphics[width=0.49\columnwidth]{sts6_stssource.pdf}%
%\caption{\CO Topography (a) and \didv\ map (b) taken on TiN. The image size is $20\times20$ nm$^2$ with $I_\mathrm{t}=300$~pA and $\Vb=520$~mV ($\Vb=273$~mV for the map). The z-scale is $\Delta_z=3.1$~nm ($z_\mathrm{RMS}=0.5$~nm) and $\Delta_{\didv}=9.8$~nS ($\didv_\mathrm{RMS}=0.2$~nS), respectively. Regions showing Coulomb blockade are marked with blue boxes. The bottom left box marks the region of interest (\ROI) chosen for this paper. \RA\ and \RB\ mark the source regions for the comparison in Fig.~\ref{cb_region_sts}.}\label{cb_region}%
%\end{figure}

\begin{figure}%
\raisebox{0.75cm}{\includegraphics[width=0.45\columnwidth]{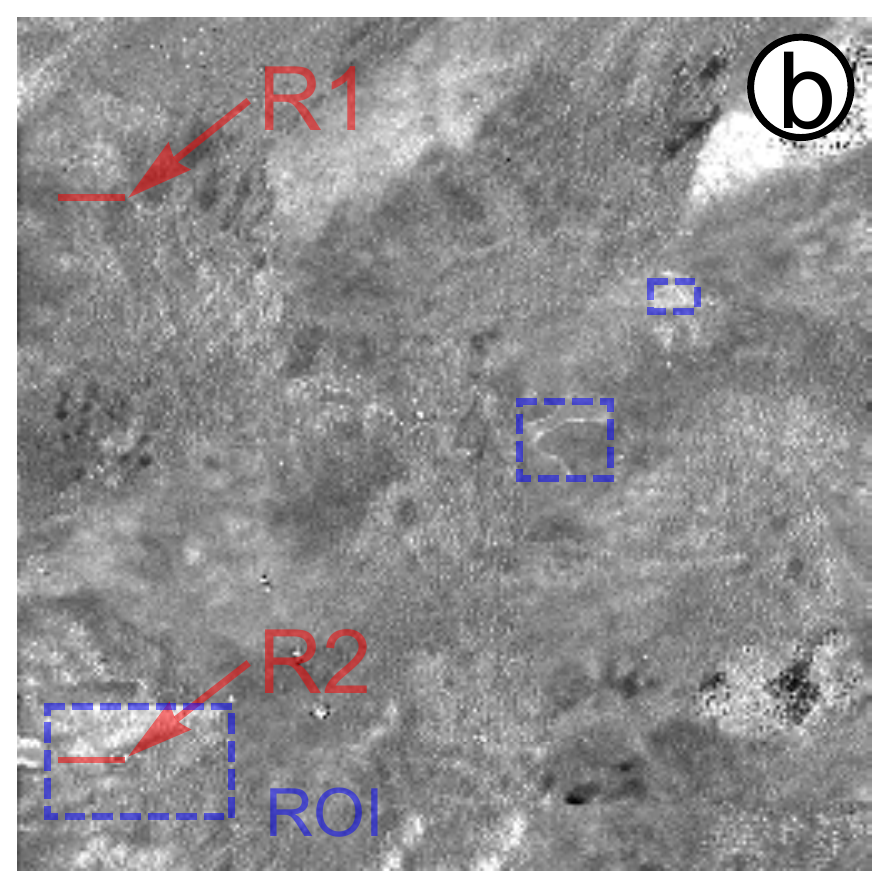}}\hfill%
\includegraphics[width=0.5\columnwidth]{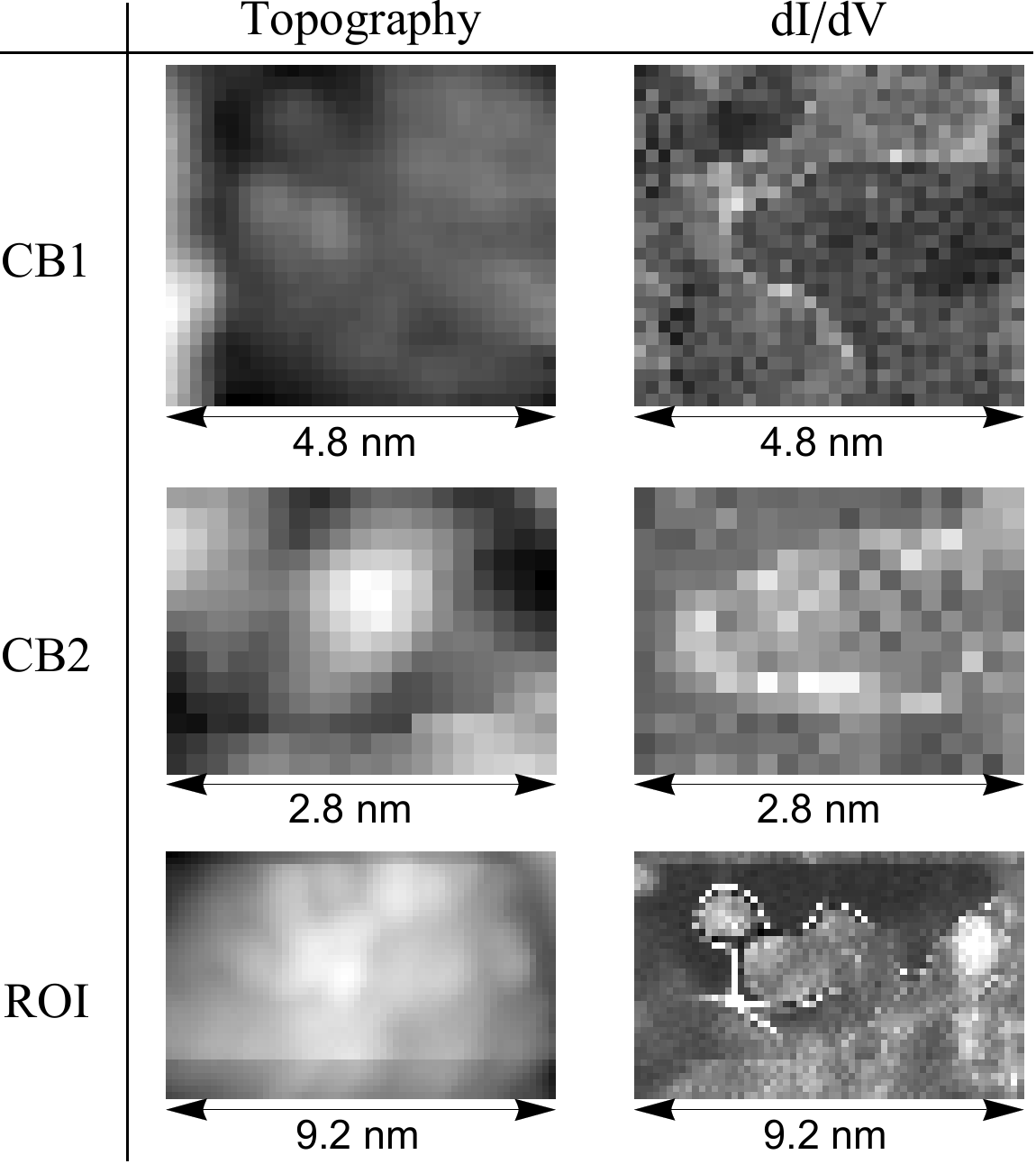}%
\caption{\CO Topography of TiN with three regions showing CB in \didv\ measurements. The image size is $40\times40$ nm$^2$ with $I_\mathrm{t}=300$~pA and $\Vb=520$~mV. The z-scale is $\Delta_z=4.7$~nm ($z_\mathrm{RMS}=0.75$~nm). Regions showing Coulomb blockade are marked with blue boxes. The bottom left box marks the region of interest (\ROI) chosen for this paper. \RA\ and \RB\ mark the source regions for the comparison in Fig.~\ref{cb_region_sts}. The table shows the local topography as well as conductance data for the three CB regions at a bias voltage of $-280$~mV, $253$~mV, and $520$~mV, respectively. The sharp line structures in the conductance maps indicate CB. }\label{cb_region}%
\end{figure}

\begin{figure*}%
\includegraphics[width=\textwidth]{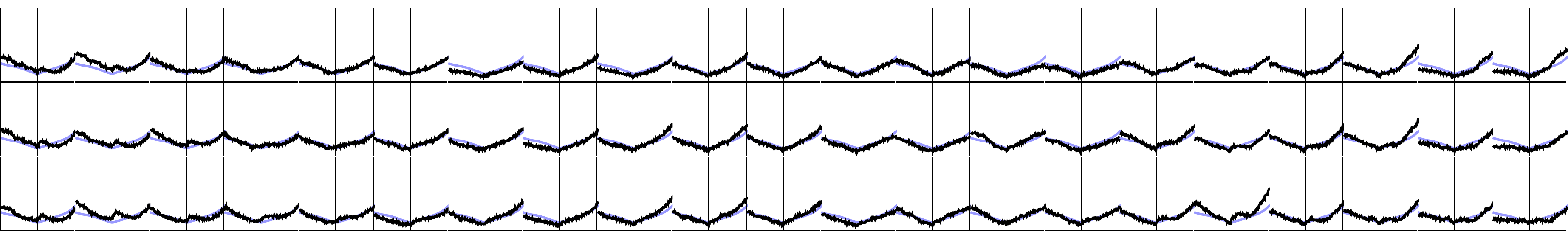}\\%
\includegraphics[width=\textwidth]{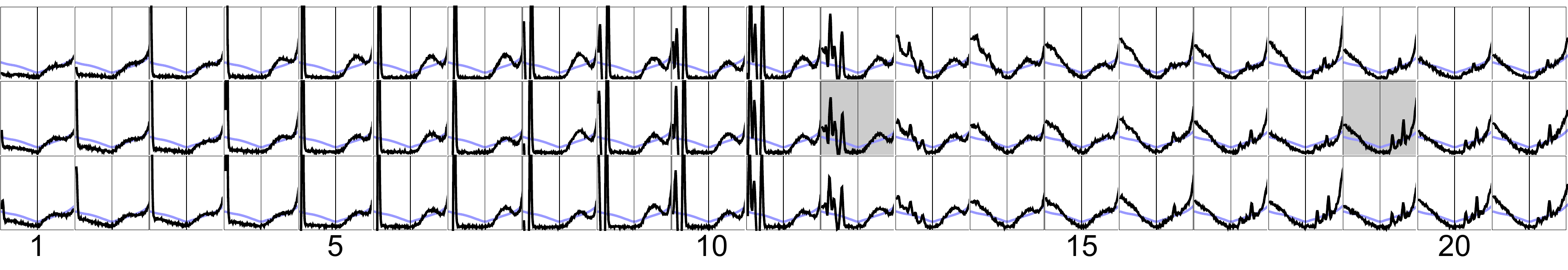}%
\caption{\CO Comparing \didv\ curves from \RA\ and \RB\ marked in Fig.~\ref{cb_region}. The plot range of each curve is $\Vb\in\left[-520,480\right]$ mV and $\didv\in\left[0,3\right]$ nS, respectively. The lateral spacing of the curves is $d=1.6$ \AA. In \RA\ the curves largely resemble the average spectrum (blue curve). In \RB\ the STS curves show a strong peak structure. The two marked spectra are further discussed below.}\label{cb_region_sts}%
\end{figure*}

Fig.~\ref{cb_region} shows the topography ($40\times40$ nm$^2$, $I_\mathrm{t}=300$~pA, $\Vb=520$~mV) and three parts of a simultaneously obtained conductance map (\ROI: $\Vb=-280$~mV, \CB1: $\Vb=252$~mV, \CB2: $\Vb=520$~mV) which show signs of CB. The areas are marked by blue  boxes in the topographic image. The \didv\ maps show sharp bright lines whose position shift with bias voltage indicating CB. For the remainder of this paper we focus on the largest region (marked \ROI). The size of this region is $8.0\times4.2$~nm$^2$ and contains $51\times27=1377$ individual \didv\ spectra.

To eliminate the possibility that the observed CB in \ROI\ stems from a particle on the tip we compare the spectroscopy of similar regions on two different crystallites. Fig.~\ref{cb_region_sts} shows $21\times3$ arrays of individual spectra from the region \RA\ and \RB\ marked in Fig.~\ref{cb_region}. A blue line in each of the graphs marks the averaged spectrum of the whole scan area (Fig.~\ref{lrg_sts} (b)). The spectra from \RA\ are virtually identical to the average. In contrast, spectra from \RB\ --- as well as most spectra in \ROI\ --- show a distinct peak structure. The peaks appear in groups with fixed peak-peak spacing within the group. The energetic position of the group as well as the peak heights vary systematically with the tip position. As the tip moves from left to right the group of peaks enters at negative \Vb\ and keep moving towards 0 V while diminishing in height. At frame 15 from the left the peaks are no longer visible and reappear at positive \Vb\ as the tip continues to move to the right.

\begin{figure}%
\includegraphics[width=0.45\columnwidth]{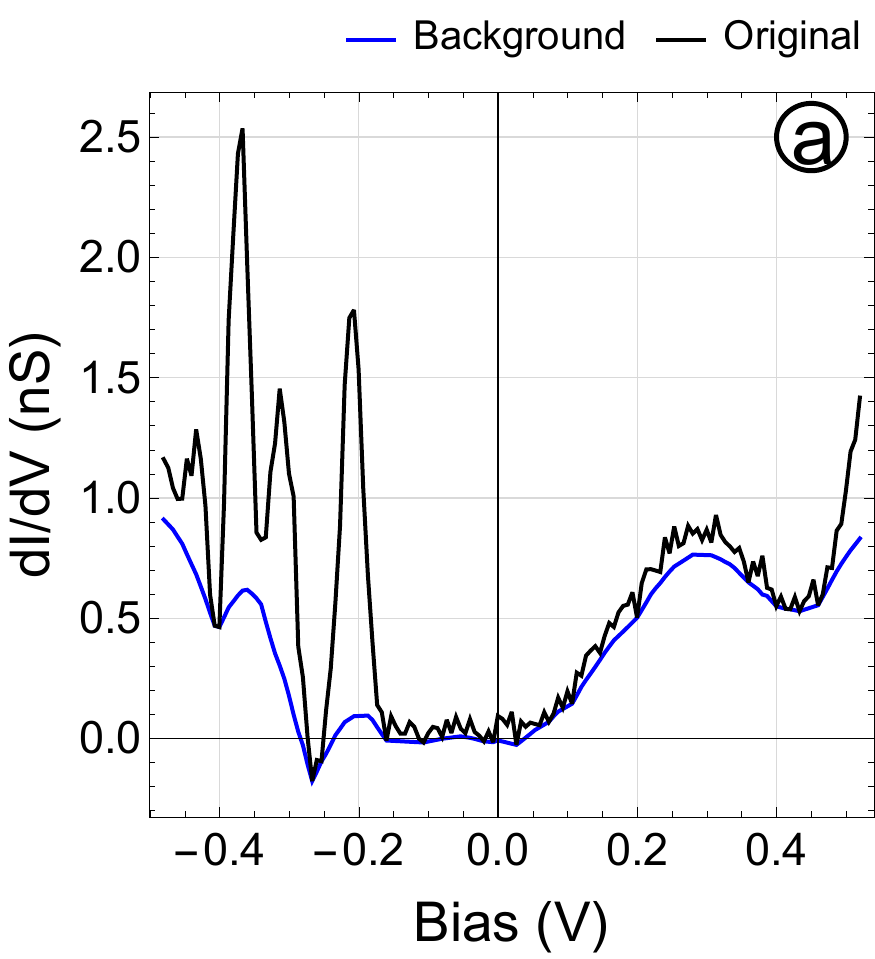}\hfill%
\includegraphics[width=0.45\columnwidth]{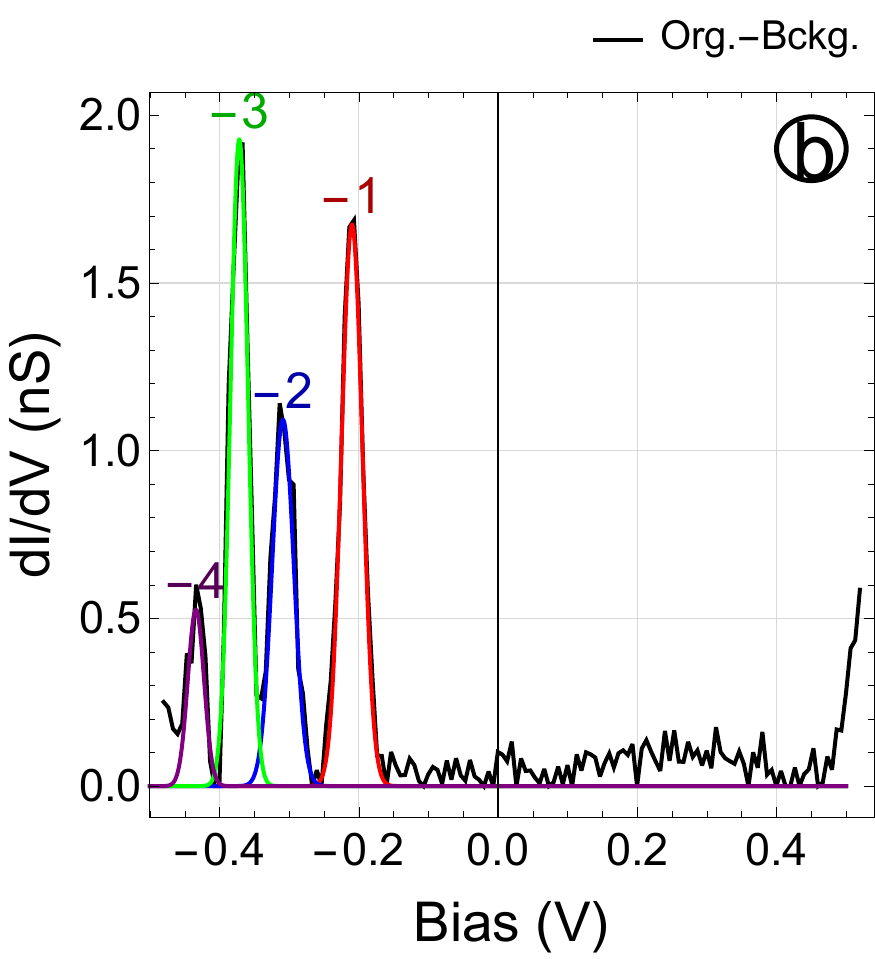}\\%
\includegraphics[width=0.45\columnwidth]{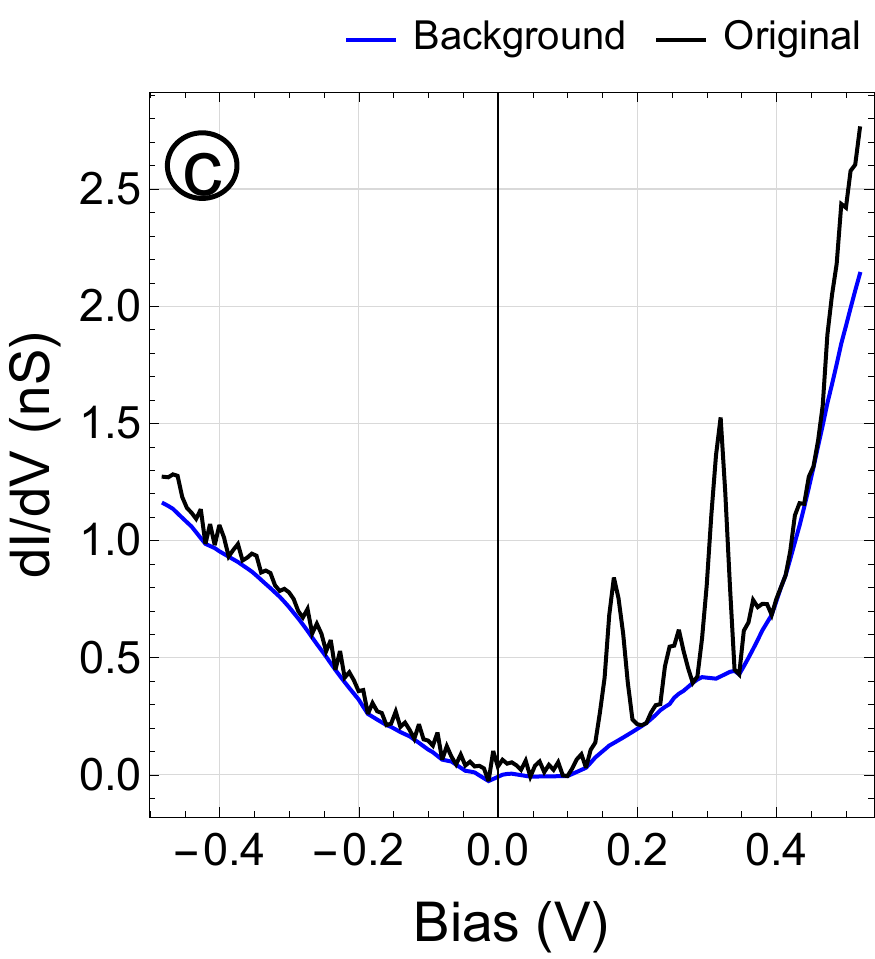}\hfill%
\includegraphics[width=0.45\columnwidth]{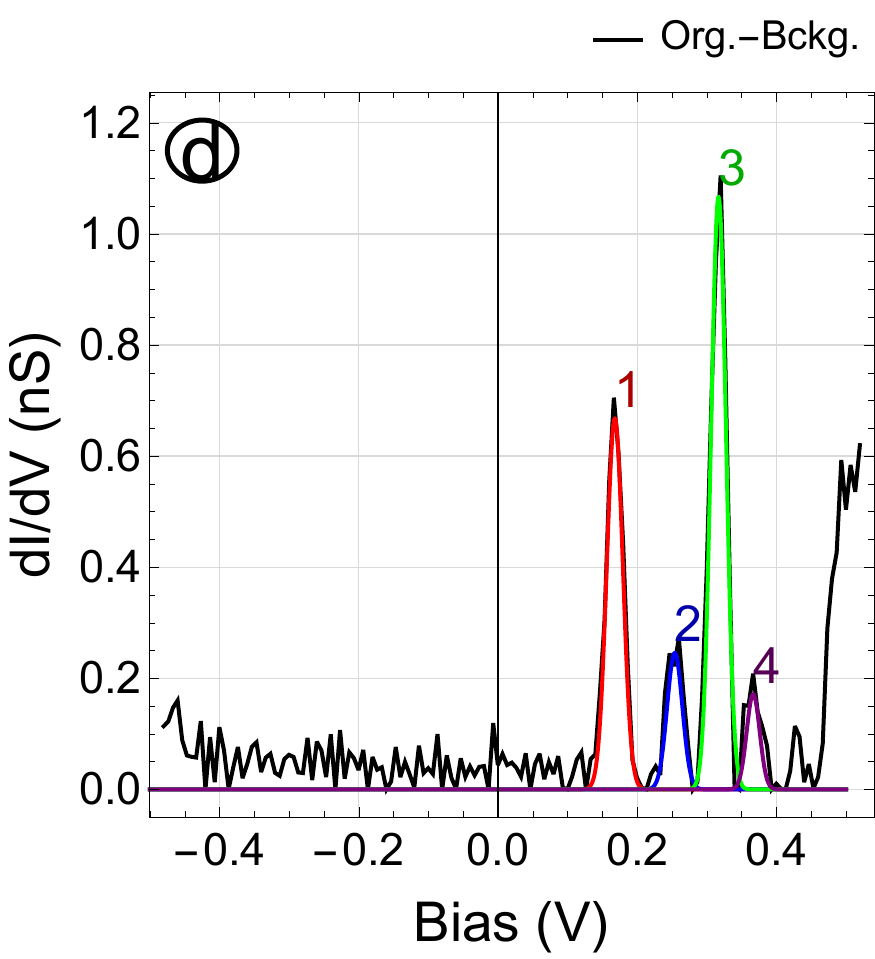}%
\caption{\CO Example spectra showing at least four peaks at a negative/positive sample bias, respectively. To extract peak position, height and width we subtracted the background and fit the individual peaks with Gaussian functions. The peaks are numbered with negative/positive numbers starting with the peak closest to 0 V as indicated.}\label{cb_get_peak}%
\end{figure}

For further analysis each of the curves in \ROI\ was semi-automatically searched for discernible peaks. An example of the process for two spectra marked in Fig.~\ref{cb_region_sts} is shown in Fig.~\ref{cb_get_peak}. Plot (a) and (c) show the original data with an estimated background. The background is then subtracted from the \didv\ data leaving sharp peaks largely unchanged as shown in plot (b) and (d). The peaks are then fit by a Gaussian function $y(V)=a\cdot\exp\left(-(V-V_0)^2/\sigma^2\right)$ to determine position $V_0$, amplitude $a$ and width $\sigma$. The peaks are numbered with negative/positive numbers starting with the peak closest to $\Vb=0$~V. The algorithm found a total of 3359 peaks in 1165 ($\sim85\%$) of the spectra of which 908 ($\sim66\%$) show at least 2 peaks. A small fraction of the automatically detected peaks are erroneous.

\begin{figure}%
\begin{tabular}{l}
\includegraphics[width=0.77\columnwidth]{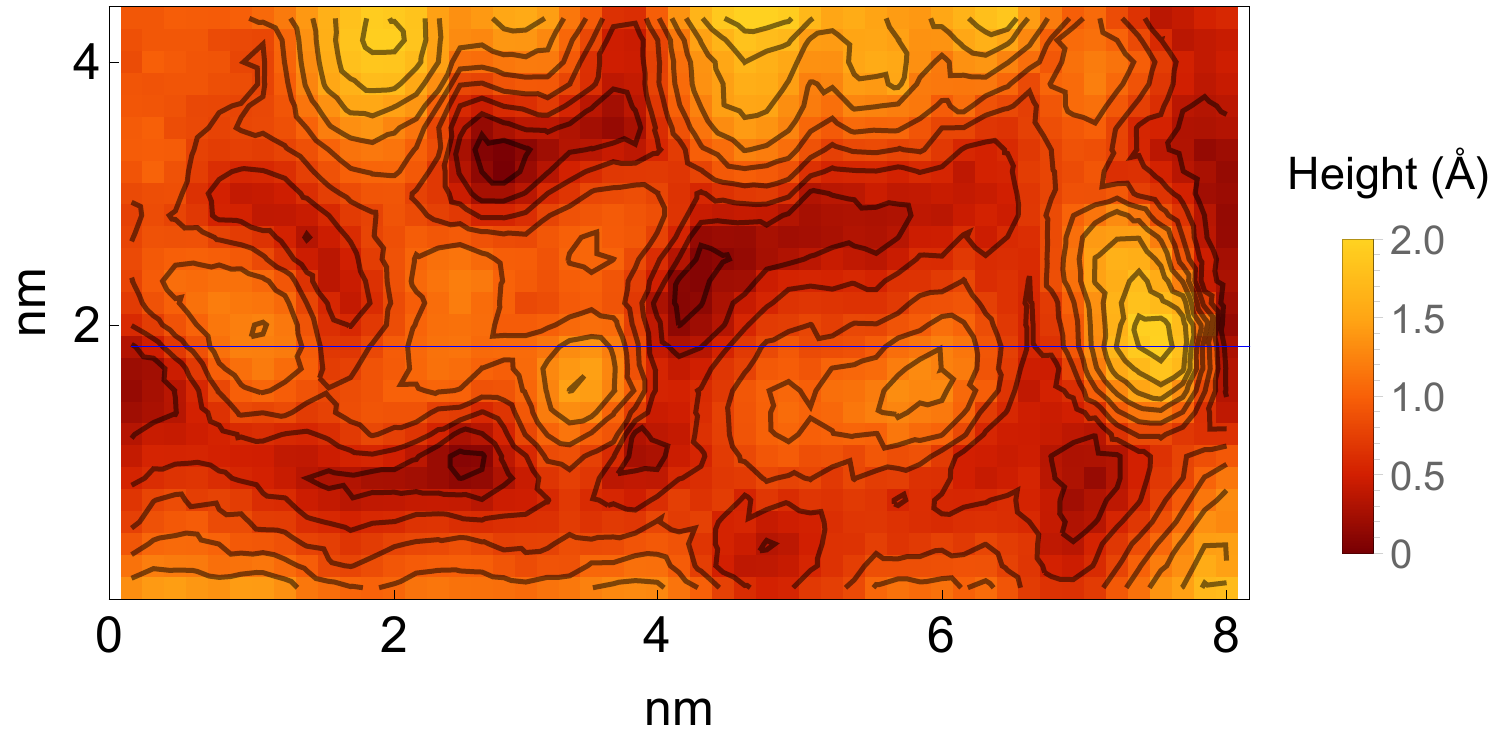}\\%
\includegraphics[width=0.95\columnwidth]{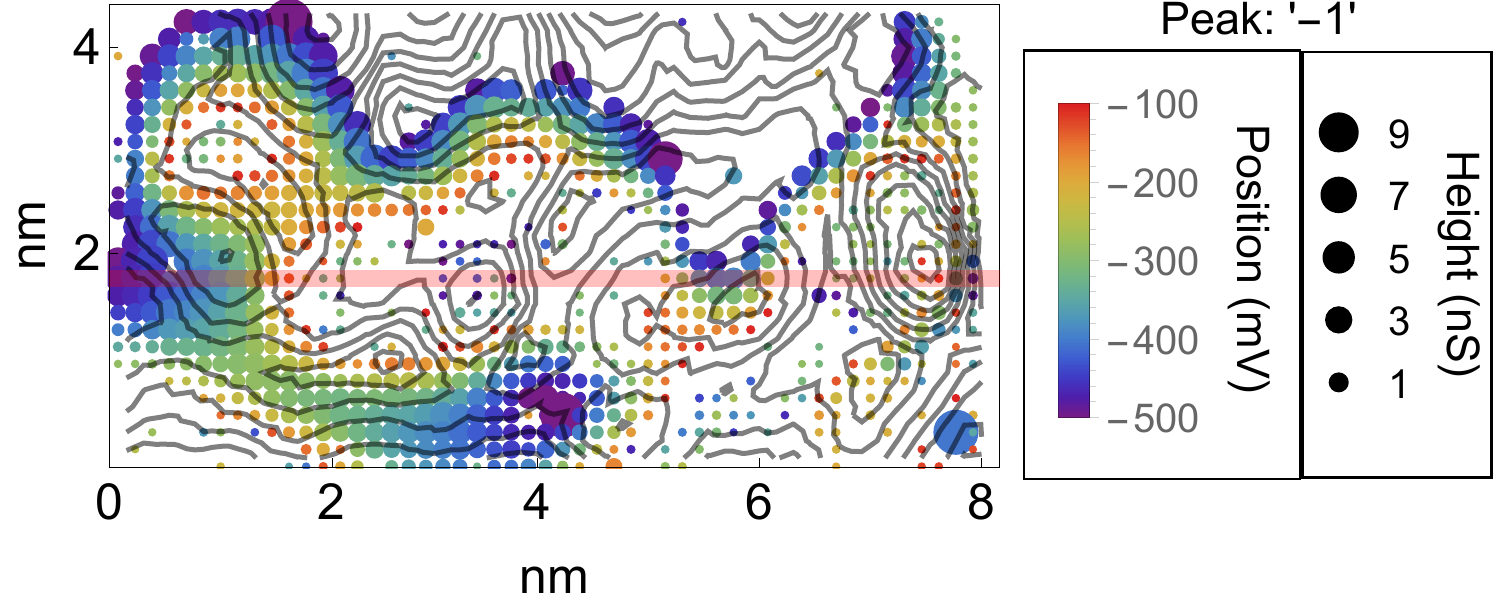}\\%
\includegraphics[width=0.95\columnwidth]{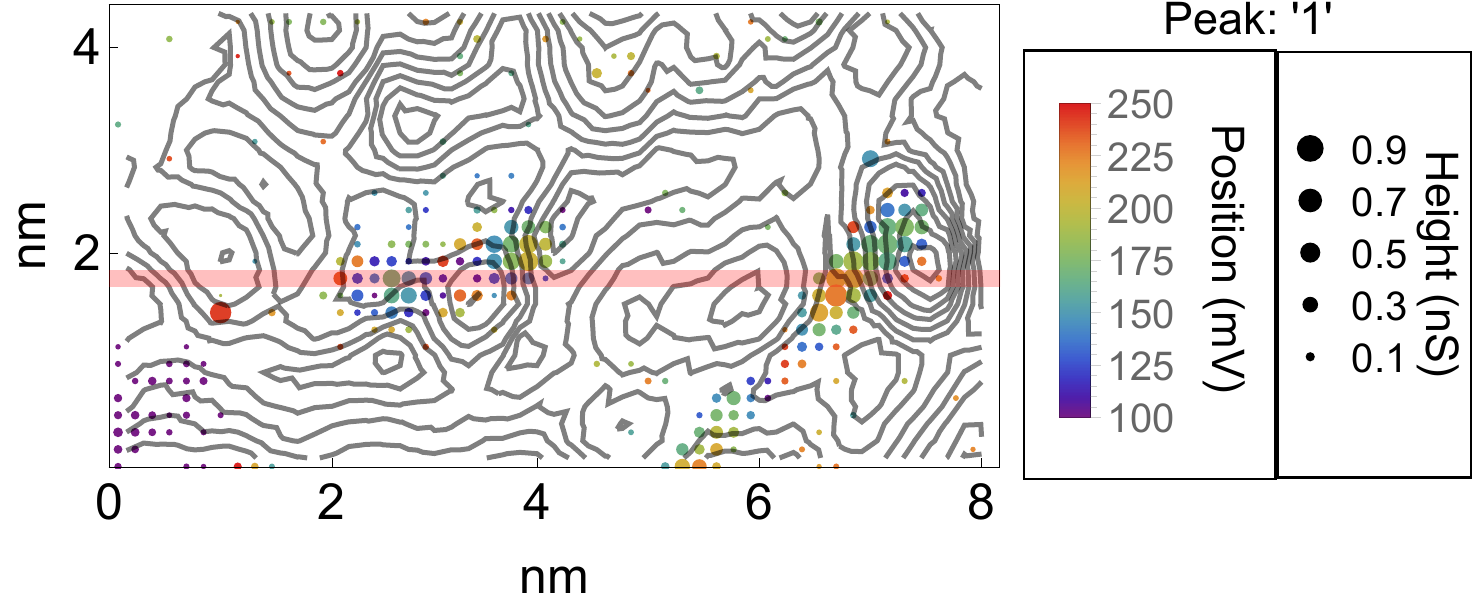}%
\end{tabular}
\caption{\CO Top to bottom: Topography of \ROI, contour plot of the topography with color coded position and height of peak ``-1'' and contour plot of the topography with color coded position and height of peak ``1''. The red shading marks the source region for Fig.~\ref{cb_firstpeak}.}\label{cb_peak1_lat}%
\end{figure}

As mentioned before, one of the signatures of CB in STM measurements is a systematic change of peak position with capacitive coupling to the tip and thereby the tip position. Such a dependance is demonstrated in Fig.~\ref{cb_peak1_lat}. It shows the evolution of peak ``-1'' and ``1'' in two dimensions. The peak position and height is represented by color and size of dots plotted on top of a contour plot of the topography. Since the peaks are not found in all spectra within this area a portion of the positions are left blank. In at least two core regions peak ``-1'' shifts towards lower energies as the tip approaches the respective center. Peak ``-1'' then vanishes and peak ``1'' appears. The latter shifts towards higher energies as the tip moves closer to the center. No correlation with topography is apparent. The details of the shift in peak position depend strongly on changes in the tip-sample capacitance and therefore on both topography and tip shape. 

\begin{figure}%
\includegraphics[width=0.9\columnwidth]{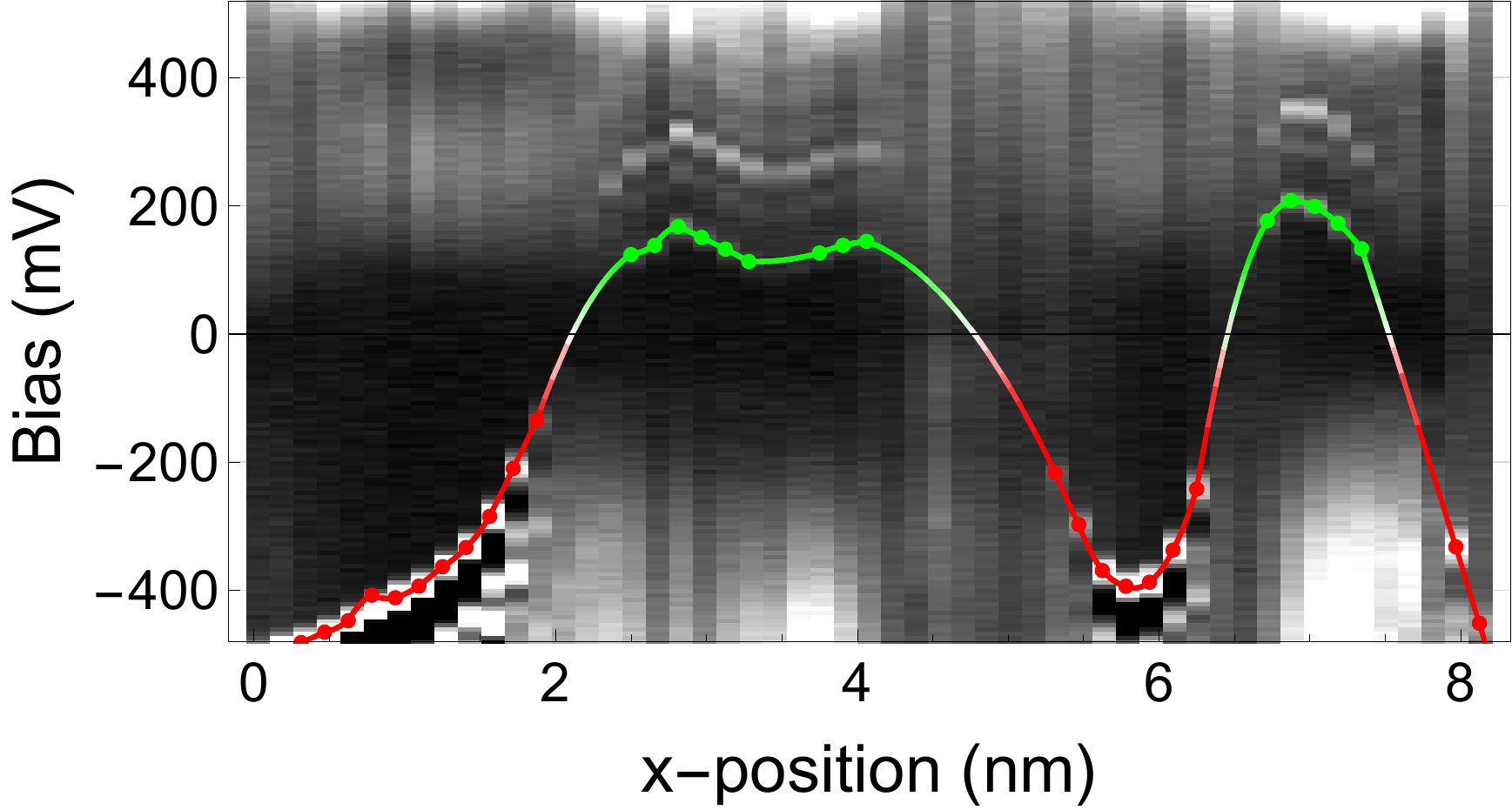}%
\caption{\CO Evolution of \didv\ spectroscopy with position along line in Fig.~\ref{cb_peak1_lat} through the whole \ROI. The gray scale image represents the measured \didv\ values as a function of position and \Vb, respectively. The fitted peak position for ``-1'' (``1'') are marked by red (green) dots. The points are connected by a spline interpolation as a guide to the eye.}\label{cb_firstpeak}%
\end{figure}

Fig.~\ref{cb_firstpeak} shows the evolution of the \didv\ spectra as the tip moves along the horizontal axis (shaded line in Fig.~\ref{cb_peak1_lat}) intersecting two center regions at which peak ``1'' appears. The automatically found peak positions ``-1'' and ``1'' are marked by red and green dots, respectively. The dots are connected by a spline interpolation as a guide to the eye. It is clear that the peak groups exclusively appear at positive or negative bias voltages. This is an unusual behavior for CB but similar to published work on dioctyldithiophosphate sodium salt molecules\cite{KLUSEK1999262}. There it was a hallmark of two vastly different junction resistances and capacitances in which case the junction with the higher resistance (lower tunneling rate) dominates. Here we assume an analogue situation as shown by simulations presented in the next section.

\begin{figure}%
\includegraphics[width=0.49\columnwidth]{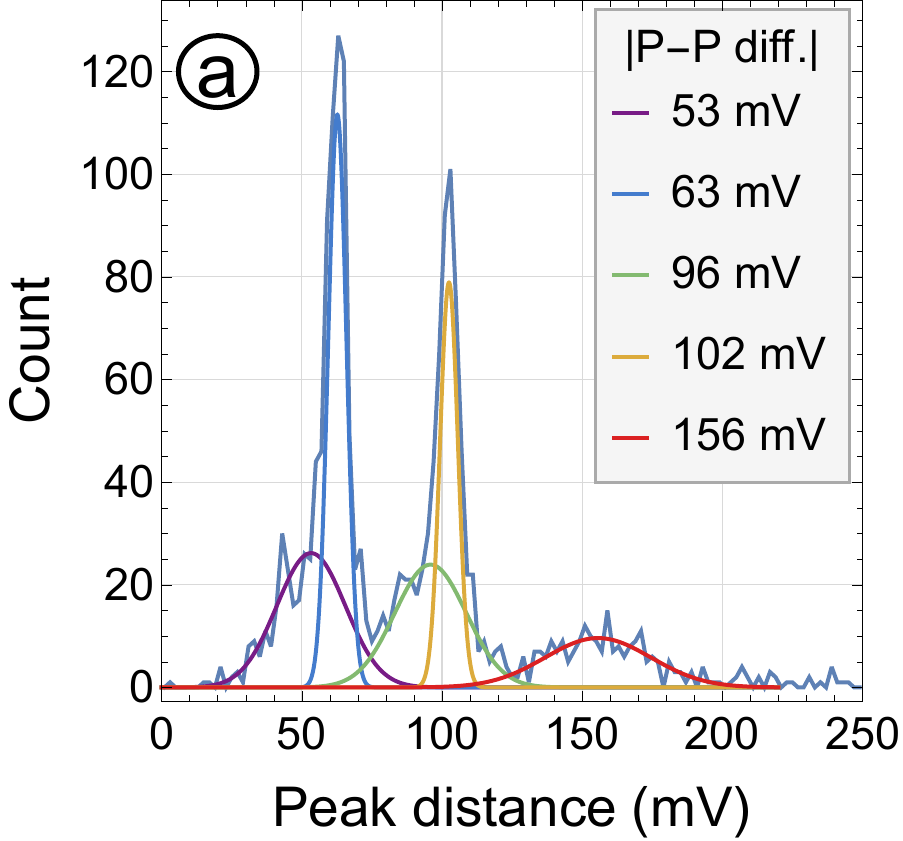}\hfill%
\includegraphics[width=0.49\columnwidth]{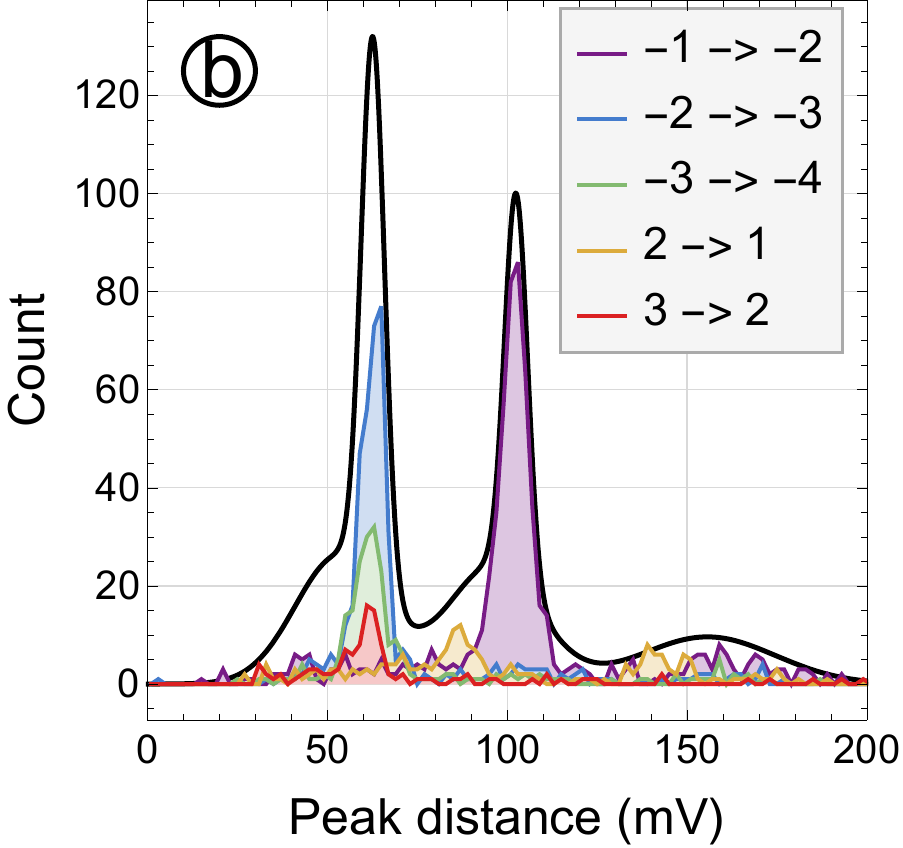}\hfill%
\includegraphics[width=0.49\columnwidth]{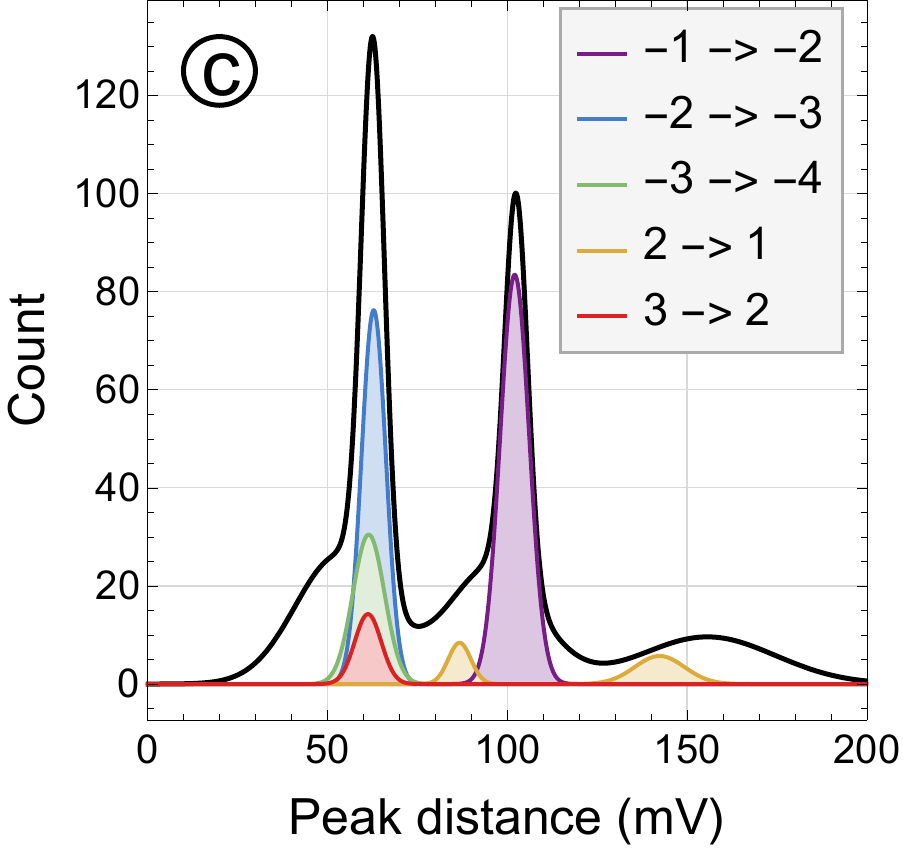}%
\caption{\CO (a) Histogram of nearest neighbor peak distances showing the principal energy scales. (b) Histograms considering peak numbers. (c) Gaussian fits to the main histogram peaks in (b).}\label{cb_peak_dist}%
\end{figure}

\begin{table}[tb]
\begin{tabular}{cccc}
\hline
Peaks&Count&Position (mV)&$\sigma$ (mV)\\
\hline
$-1\to-2$&425&102&4.0\\
\hline
$-2\to-3$&327&63&3.4\\
\hline
$-3\to-4$&166&62&4.4\\
\hline
$1\to2$&33&87&3.1\\
``$1\to2$'' $(1\to3)$&50&142&6.8\\
\hline
$2\to3$ $(3\to4)$&66&62&3.6\\
\hline
\end{tabular}
\caption{Average peak distances and number of counts under the respective curve in the histogram shown in Fig.~\ref{cb_peak_dist}. The $1\to3$ count occurs when the small ``2'' peak is missed by the fit procedure. Consequently, some of the $2\to3$ counts contain $3\to4$ distances which are similar in value.}\label{cb_peak_distT}
\end{table}

Finally, we analyze the distance between neighboring peaks within the \didv\ spectra. Fig.~\ref{cb_peak_dist} (a) shows a histogram of all distances found in \ROI. The resulting distribution is well approximated by five Gaussian functions. However, an analysis considering the origin of the distance --- i.e. the peak numbers involved --- gives a more refined picture. Fig.~\ref{cb_peak_dist} (b) shows histograms of the five most common pairs. For comparison the outline of the overall histogram is also shown. Again, the histogram peaks (HP) were fit by Gaussians and the result is shown in Fig.~\ref{cb_peak_dist} (c) and summarized in Table \ref{cb_peak_distT}. The HP positions fall into two groups: the innermost ones ($V_{-1\to-2}=102$~meV and $V_{1\to2}=87$~meV) and the rest around $V_\mathrm{p}\approx 62$~meV. The additional HP of $V_{``1\to2``}=142$~meV stems mostly from missing the relatively small peak ``2'' in our automatic peak detection and thus measuring the $1\to3$ distance instead. The quantum dot states show an identical level spacing when appearing on the positive/negative bias side of $V_\mathrm{p}\approx 62$~meV except for the first distances. While the QD is geometrically defined the ground state might depend on the bias polarity or QD charging state. As we will show in the next section the switch of the peaks from negative to positive sample bias is indeed accompanied by the addition of one electron to the QD. This could explain the difference between $V_{2\to1}$ and $V_{-1\to-2}$.

\subsection{Simulations}

\begin{figure}%
\includegraphics[width=0.8\columnwidth]{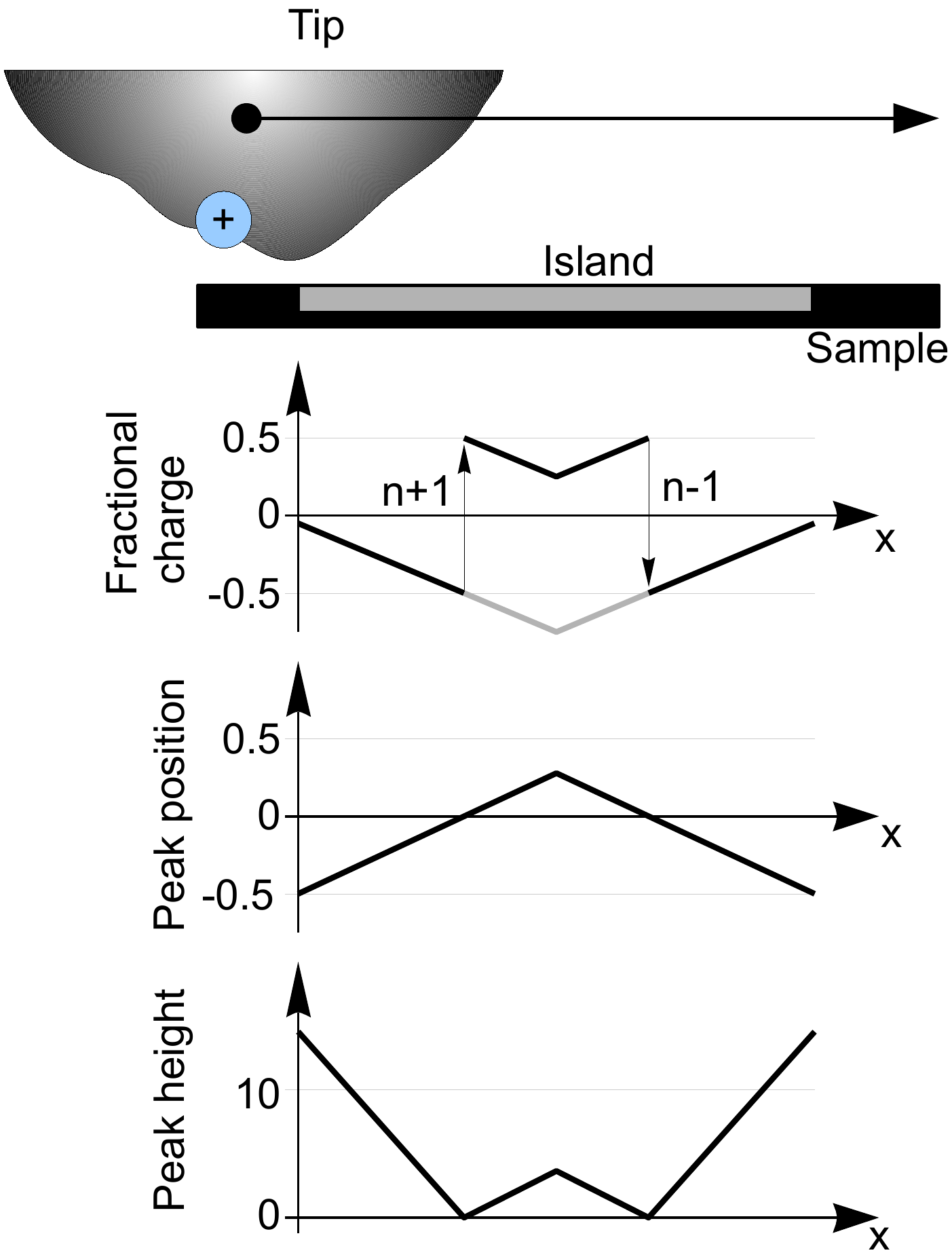}%
\caption{Model for the theoretical calculations. As the tip carrying a positive charge moves across the CB island it changes the fractional charge \Qf\ on the island from $\Qf\sim 0 \qe$ towards $\Qf=-0.5 \qe$. This shift moves the position of the lead peak of the CB from negative bias voltages towards zero. At the same time the peak height diminishes. Reducing \Qf\ beyond $-0.5 \qe$ leads to a charging event of the island by adding one electron after which \Qf\ changes from $0.5 \qe$ towards zero. During this time the position of the leading CB peak moves from zero to more positive bias voltages while its height increases. This trend continues until the tip reaches the (electrical) center of the island and reverses as the tip approaches the opposite edge of the CB region.}\label{cb_theory_model}%
\end{figure}

\begin{figure}%
\includegraphics[width=0.95\columnwidth]{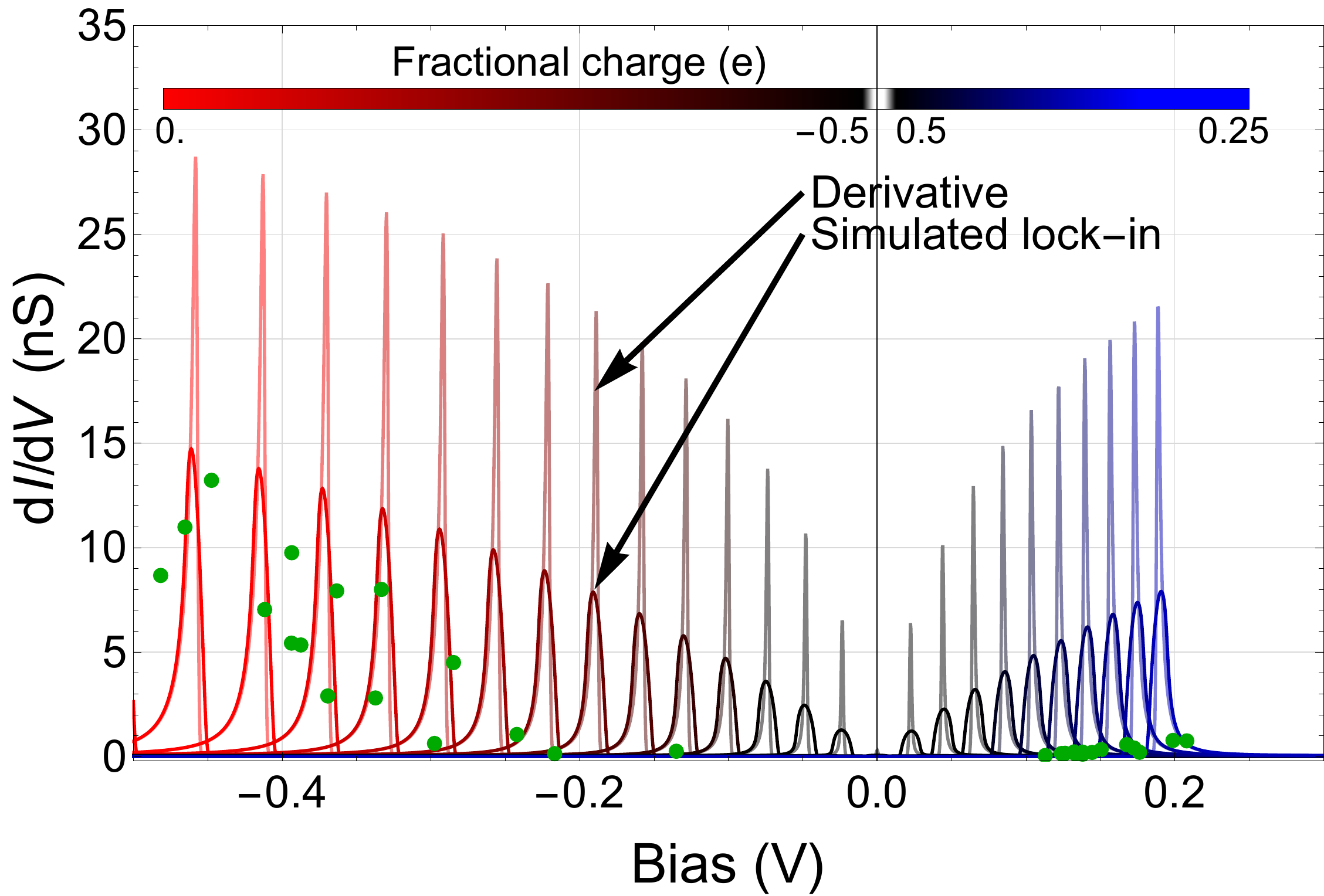}%
\caption{Comparison of charging peaks according to orthodox theory (curves) and measured peak position vs.\ peak height (green dots). The curves represent 24 calculations showing one CB peak each. The peak position moves towards higher energy with decreasing \Qf. The tip--sample capacitance is simultaneously increased which accounts for the change in peak--peak spacing. The parameters for this calculation are summarized in Tab.\ \ref{cb_peak_pars}. The \didv\ curves are calculated from the simulated $I(V)$ curves by directly calculating the derivative and by simulating the response of a lock in amplifier.}\label{cb_theory}%
\end{figure}

\begin{table}[tb]
\begin{tabular}{|c|c|c|}
\hline
Parameter&STM junction (1)&Sample junction (2)\\
\hline
Resistance&1.67 G$\Omega$&20 M$\Omega$\\
Capacitance&0.1--0.2 aF&1 zF\\
\hline
\Qf&\multicolumn{2}{c|}{$\left[-0.45,0.45\right]\qe$}\\
Temperature&\multicolumn{2}{c|}{4.2 K}\\
\hline
\end{tabular}
\caption{Parameters for CB simulation. }\label{cb_peak_pars}
\end{table}

Previous observations of Coulomb blockade using STM were performed mostly on special samples of small metallic particles at room temperature\cite{C4CS00204K,PhysRevB.61.10929,PhysRevB.48.12104} or low temperature\cite{FrontPhys.1.13,doi:10.1063/1.3624952,Miller1994,SCHONENBERGER1993222,KLUSEK1999262,B605436F,PhysRevB.95.081409}. All of these observations are usually explained by orthodox Coulomb blockade theory\cite{PhysRevB.43.1146,PhysRevB.44.5919} which calculates the tunnel rates between two electrodes and an island and thus the resulting current flow between the electrodes. These calculations allow the estimation of the resistance and capacitance of the two junctions as well as the fractional electron charge on the island. The absence of a dedicated gate electrode limits the analysis compared to a regular SET circuit.

In our data it is clear that while the peak groups shift with tip position the distances within the group remain constant. This is consistent with tunneling through a quantum dot with internal states. A detailed simulation such as given for a semiconducting QD in\cite{B605436F} is beyond the scope of this paper. However, some of the behavior can be explained using orthodox CB theory. We only consider the position and height of the innermost peaks (``-1'' and ``1''). The energy of peak $n$ can be expressed analytically by the junction capacitance $C_1$ and $C_2$ and the fractional charge on the island \Qf\ by
\[ E_\mathrm{1,n}=\qe\cdot\frac{\pm\frac{1}{2}-\Qf+n}{C_1} \]
and
\[ E_\mathrm{2,n}=-\qe\cdot\frac{\pm\frac{1}{2}-\Qf+n}{C_2} \]
Here, however, we only observe a single CB peak in each spectrum which leaves no practical way to decouple the parameters $n$, $C_1$, $C_2$ and \Qf. Therefore we use the correlation between peak height and position as a function of \Qf\ as the principle driver. Fig. \ref{cb_theory_model} shows the proposed behavior. We assume that the tip carries a positive charge conceivably in remaining oxide layers near the apex. As the tip moves across the CB island it induces an increasing negative fractional charge \Qf\ on the island from $\Qf\sim 0 \qe$ towards $\Qf=-0.5 \qe$. This change in \Qf\ moves the position of the lead peak of the CB from negative bias voltages towards zero. At the same time the peak height diminishes. Reducing \Qf\ beyond $-0.5 \qe$ leads to a charging event of the island by adding one electron. This change in charge might also be responsible for the difference in the peak $-1\to-2$ and $1\to2$ distance we observed. As the tip proceeds to move towards the (electronic) center of the island  \Qf\ decreases from now $0.5 \qe$ towards zero. This shifts the lead CB peak towards positive voltages while its height grows again. The sequence reverses as the tip moves past the center of the island. 

We applied two methods to compute the conductance (\didv) from the \iv\ curves calculated using orthodox theory. Numerical differentiation leads to sharp peaks only thermally broadened by $T=4.2$~K. The second method simulates a lock-in amplifier using our modulation amplitude of $V_\mathrm{mod}=7$~mV. This results in broader and shallower peaks which decrease faster in height. Fig. \ref{cb_theory} compares the result of the simulation with the observed peak height vs. position. The parameters for the simulation are summarized in table \ref{cb_peak_pars}. The resistance of the STM junction was taken from the bias voltage and current set point of $\Vb=500$~mV and $I_\mathrm{SP}=0.3$~nA ($R_1=1.67$~G$\Omega$). A sample junction of $R_2=20$~M$\Omega$ then sets the overall height of the conductance peaks to roughly match the observed values of up to 15~nS. The sample side capacitance $C_2$ as well as the STM junction capacitance $C_1$ value were chosen to be small enough so that only a single peak appears within the STS window of $\Vb=\left[-500,500\right]$~mV. $C_1$ was also allowed to increase as the tip approaches the center which accounts for the increase in peak density at positive bias voltages in the simulation. Finally, $Q_f$ is varied from -.05\qe\ to -.45\qe\ to shift the peak from -0.5 V towards 0 V. This also reduces its height (red curves in Fig.~\ref{cb_theory}). Further reduction in $Q_f$ leads to a charging event changing the number of electrons $n$ by one (which is unimportant in this model), switching $Q_f$ to 0.45\qe\ and moving the peak to the positive bias side. As $Q_f$ continues to decrease, the peak shifts towards more positive bias voltages and it's height increases again (blue curves in Fig.~\ref{cb_theory}. Green dots mark the measured peak height vs.\ position. While the overall trend is well reproduced the peak heights diminish faster in our measurements than in the simulation. This discrepancy could be due to the quantum dot nature of the Coulomb blockade island or due to the background density of states of the TiN sample which was not considered in the simulation.

%CB in graphine sheet \cite{PhysRevB.95.081409}

%why are -1->-2 ne 1->2

%CB common, QD not common

\section{Conclusion}

We performed an STM study at 4.2~K of a polycrystalline TiN film with $d\approx 19$~nm diameter crystallites. STS measurements taken over a few hundred mV show a pseudo-gap at $\Vb=0$~V, similar to theoretical DFT calculations near Mott insulation. In certain locations we observed Coulomb blockade peaks as well as quantum dot states. The behavior of the Coulomb blockade peaks were explained by orthodox theory. The fractional charge on the CB island changes with tip position which in turn moves the energetic position and height of the CB peaks. While the energy spacing is too large to directly interfere with typical microwave devices operating in the GHz regime ($1$~$\mathrm{GHz}\sim 4.1$~$\mu\mathrm{eV}$), both the quantum dot and Coulomb blockade can act as a source or a trap of quasiparticles. Since CB regions were found on all samples their density has to be considerable in films prepared under similar conditions (here 1 in 266~nm$^2$). If these anomalous CB features are generic to TiN films, it would be interesting to explore the effects on device performance made by a gate which can change the charging state of the CB ensemble.

\appendix

\section{Mathematica code}
\subsection{Automatic peak detection}
The Mathematica code for the automatic peak detection in the \didv\ spectra is shown below. \texttt{ddsb} contains all \didv\ curves with the background removed using the Mathematica function \texttt{EstimatedBackground} with a scale parameter of 5. \texttt{vlt} contains a list of the voltages corresponding to the points on the \didv\ curves.
\lstset{basicstyle=\tiny,breaklines=true,mathescape=true,escapechar=\%,commentstyle=\textit}
\begin{lstlisting}
(* keep the result *)
plist =
(* need for speed *)
 Parallelize[
(* consider all elements *)
  Table[        
(* Gaussian kernel to smooth data for initial peak search *)
   kern = Normalize[Exp[-(#)^2/2/.010^2]& /@ vlt$[\![$70;;82$]\!]$]/2.; 
(* calculate half of the kernel length *)
   klh = Floor[(Length[kern]+1)/2];
(* pick one curve with background subtracted out of data set *)
   dsorg = ddsb$[\![$All,yy,xx$]\!]$;
(* smooth curve, cut of ends *)
   ds = ListConvolve[kern,dsorg,{1,-1},0]$[\![$klh;;-klh$]\!]$;
(* use FindPeaks to get initial position estimate of peak position in single curve, return {vlt,height} *)
   p = {vlt$[\![$#$[\![$1$]\!]]\!]$,#$[\![$2$]\!]$}& /@ FindPeaks[ds,.05,{.003,.05},{.09,.05}];
(* Build fit function with one Gaussian per peak *)
   fitfun = Plus @@ Table[gpf[a[n],$\mu$[n],$\sigma$[n],x],{n,Length[p]}]+y0;
(* Build parameter array for fit function *)
   pars = Join[Flatten[Table[{{a[n],p$[\![$n,2$]\!]$},{$\mu$[n],p$[\![$n,1$]\!]$},{$\sigma$[n],0.005}},{n,Length[p]}],1], {{y0,0.005}}];
(* First fit against filtered data set ...*)
   fitp = NonlinearModelFit[Transpose[{vlt,ds}],fitfun,pars,x, MaxIterations->10000] // Quiet;
(* ... then use the result as starting parameters to refine against unfiltered data *)
   fitp2 = NonlinearModelFit[Transpose[{vlt,dsorg}],fitfun,{#$[\![$1$]\!]$,#$[\![$2$]\!]$}& /@ fitp["BestFitParameters"],x,MaxIterations->10000] // Quiet;
(* List of all output parameters, include position and peak number *)
   out = Table[{$\mu$[n],a[n],$\sigma$[n],xx,yy,0}, {n,Length[p]}];
(* Fill in fit parameters parameters from first fit *)
   all = out /. fitp2["BestFitParameters"];
(* Keep only good fits $\sigma$<50 mV, 80.5>height>0, $\mu$<500 mV *)
   res = Select[all, (Abs[#$[\![$3$]\!]$]<.05 && #$[\![$2$]\!]$<80.5 && #$[\![$2$]\!]$>0 && #$[\![$1$]\!]$<.5)&];
(* All peaks on the right ... *)
   p = Flatten[Position[res$[\![$All,1$]\!]$,n_/;n>=0]];
(* ... set peak number *)
   If[Length[p]>0, oo = p$[\![$Ordering[res$[\![$p,1$]\!]]\!]$];(res$[\![$oo$[\![$#$]\!]$,6$]\!]$=#)& /@ Range[Length[p]]];
(* All peaks on the left ... *)
   p = Flatten[Position[res$[\![$All,1$]\!]$,n_/;n<0]];
(* ... set the peak numbers *)
   If[Length[p]>0, oo = p$[\![$Ordering[res$[\![$p,1$]\!]$]$]\!]$;(res$[\![$oo$[\![$#$]\!]$,6$]\!]$=#)& /@ -Range[Length[p]]];
(* put result in the table *)
   res,
(* For all curves in the region of interest *)
   {yy,207,233},{xx,15,65}]];
\end{lstlisting}

\subsection{Orthodox Coulomb blockade}
Mathematica code for calculating the Coulomb blockade tunneling current.
\begin{lstlisting}
(*total capacitance*)
CS=(C1+C2);
(*charging energies (equation (2)%\cite{PhysRevB.44.5919}%)*)
dE1p[n_,V_]=e/CS*(C2*V+e*(n-Qf)+e/2);
dE1m[n_,V_]=e/CS*(-C2*V-e*(n-Qf)+e/2);
dE2p[n_,V_]=e/CS*(-C1*V+e*(n-Qf)+e/2);
dE2m[n_,V_]=e/CS*(C1*V-e*(n-Qf)+e/2);
(*tunnel rates (equation (1)%\cite{PhysRevB.44.5919}%)*)
g1p[n_,V_]=-1/e^2/R1*dE1p[n,V]/(1-Exp[dE1p[n,V]/kb/T]);
g1m[n_,V_]=-1/e^2/R1*dE1m[n,V]/(1-Exp[dE1m[n,V]/kb/T]);
g2p[n_,V_]=-1/e^2/R2*dE2p[n,V]/(1-Exp[dE2p[n,V]/kb/T]);
g2m[n_,V_]=-1/e^2/R2*dE2m[n,V]/(1-Exp[dE2m[n,V]/kb/T]);
(*helper (equation (3)%\cite{PhysRevB.43.1146}%)*)
xg[n_,V_]=g1p[n,V]+g2p[n,V];
yg[n_,V_]=g1m[n,V]+g2m[n,V];
(*density of states, cut off at $\pm$bnd (equation (5)%\cite{PhysRevB.43.1146}%)*)
$\rho$g[n_,V_,bnd_]:= 
  Product[xg[i,V],{i,-bnd,n-1}]*
   Product[yg[i,V],{i,n+1,bnd}]/
    Sum[Product[xg[i,V],{i,-bnd,j-1}]*
      Product[yg[i,V],{i,j+1,bnd}],{j,-bnd+1,bnd-1}];
(*current with cut off parameters nm, bnd (equation (3)%\cite{PhysRevB.44.5919}%)*)
Itg[V_,nm_,bnd_]:=
  e*Sum[(g2p[n,V]-g2m[n,V])*$\rho$g[n,V,bnd], {n,-nm, nm}];
\end{lstlisting}

\subsection{Calculation of dI/dV}

Using Mathematicas symbolic differentiation function D the derivative of the expression for the current with respect to the bias voltage is first calculated. The resulting formula is the used to generate the plot.

\begin{lstlisting}
(*Parameters and constants*)
vals = {e->1.6*^-19,kb->1.381*^-23,T->4.2,C1->1*^-19,C2->0.1*^-19,R1->Rmeas,R2->.2*^8,Qf->0};
(*Create table of plots with varying $Q_f$ and $C_1$*)
plts=Parallelize[Table[
(*$Q_f$ varies -.05 to -.5, .5 to .25*)
    qqf=If[r<=.5,-r,1.-r];
(*$C_1$ varies 10.075 to 11.125 aF*)
    cc1=(C1+r*15.*^-20)/.vals;
(*color scale based on $Q_f$*)
    ci=(Abs[qqf]-.18)*4;
(*Plot color*)
    col= 
     Blend[{If[qqf<0,Lighter[Red,.5], Lighter[Blue,.5]], 
       Lighter[Black,.5]}, ci];
(*Generate the plot*)
    Plot[Evaluate[
      D[Itg[V,1,3], V]*1*^9/.{Qf->qqf,C1->cc1}/. 
       vals], {V,-0.5,.5},PlotRange->All,PlotStyle->col],{r, 
     0.05,.75,0.025}]];
(*Overlay calculation and data (gdots)*)
Show[Join[{ListPlot[{0, 0}, PlotStyle->None, 
    PlotRange->{{-.5,.3},{0,30}}]}, 
  plts, {ListPlot[{#$[\![$1$]\!]$,#$[\![$2$]\!]$*1*^9}&/@ gdots$[\![$All,{1, 2}$]\!]$]}]]
\end{lstlisting}

The lock-in simulation is based on a convolution of the $I(V)$ curve with one period of a cosine function while modulating the bias voltage, i.e.:
\[\didv_\mathrm{LI}\left(V\right)=\frac{1}{a_\mathrm{LI}\pi}\cdot\int_0^{2\pi}\mathrm{d}t\cos(t)\cdot I\left(V+a_\mathrm{LI}\cdot\cos\left(t\right)\right)
\]
 
\begin{lstlisting}
(*Parameters and constants*)
vals={e->1.6*^-19,kb->1.381*^-23,T->4.2,C1->1*^-19,C2->0.1*^-19,R1->Rmeas,R2->.2*^8,Qf->0};
(*Assumed modulation amplitude in Volt*)
liamp=0.0067;
(*Create table of plots with varying $Q_f$ and $C_1$*)
plts2=Parallelize[Table[
(*$Q_f$ varies -.05 to -.5, .5 to .25*)
    qqf=If[r<=.5,-r,1.-r];
(*$C_1$ varies 10.075 to 11.125 aF*)
    cc1=(C1+r*15.*^-20)/.vals;
(*color scale based on $Q_f$*)
    ci=(Abs[qqf]-.18)*4;
(*Plot color*)
    col=Blend[{If[qqf<0,Red,Blue],Black},ci];
(*Generate the plot*)
    Plot[
     1*^9*
      NIntegrate[
       Cos[x]*Itg[V+liamp*Cos[x],1,3]/$\pi$/liamp/.{Qf->qqf, 
          C1->cc1}/.vals,{x,0,2$\pi$}],{V,-0.5,.5}, 
     PlotRange->All,PlotStyle->col],{r,0.05,.75,0.025}]];
(*Overlay calculation and data (gdots)*)
Show[Join[
  plts2,{ListPlot[{#$[\![$1$]\!]$,#$[\![$2$]\!]$*1*^9}&/@gdots$[\![$All,{1, 2}$]\!]$]}]]
\end{lstlisting}
\bibliographystyle{apsrev4-1}
\bibliography{Tin_SET}

\end{document}